# A hybrid neural network model based on improved PSO and SA for bankruptcy prediction

Fatima Zahra Azayite[1] and Said Achchab[2]

[1] National School for Computer Science and Systems analysis, Mohammed V University
Rabat, Morocco

[2] National School for Computer Science and Systems analysis, Mohammed V University
Rabat, Morocco

**Abstract**
Predicting firm's failure is one of the most interesting subjects for investors and decision makers. In this paper, a bankruptcy prediction model is proposed based on Artificial Neural networks (ANN). Taking into consideration that the choice of variables to discriminate between bankrupt and non-bankrupt firms influences significantly the model's accuracy and considering the problem of local minima, we propose a hybrid ANN based on variables selection techniques. Moreover, we evolve the convergence of Particle Swarm Optimization (PSO) by proposing a training algorithm based on an improved PSO and Simulated Annealing. A comparative performance study is reported, and the proposed hybrid model shows a high performance and convergence in the context of missing data.

## 1. Introduction

Predicting failure is one of the most challenging research topics in the financial field, especially in recent decades. It is the ability to predict if a company will fall in bankruptcy or not. It is an essential key indicator for decision-making in many cases, particularly in the areas of investment and lending. Setting up an early warning system to predict business failures based on their financial behavior can significantly improve decision-making.

Early studies started by Beaver [1] and Altman [2] were focused on designing predicting failure models to discriminate between failing and non-failing firms. These models have been implemented based on statistical techniques especially Discriminate Analysis. With the appearance of artificial intelligence, many studies have been interested to use these techniques. One of the most popular and performer intelligent tool is Artificial Neural Network (ANN) model.

ANN models discriminate between bankrupt and non-bankrupt firms based on a set of variables. So, the set of appropriate financial ratios is one of the parameters that can affect the ANN performance. In literature, there are no universally agreed financial variables to use in bankruptcy prediction models. For instance, Altman [2] used 5 ratios, Liang & Wu [3] used 7 ratios, [4] used 35 ratios. In this context, F. Du jardin [5] has pointed that a neural-network-based model for predicting bankruptcy performs significantly better when designed with appropriate variable selection techniques than when designed with variables commonly used in the financial literature. For this purpose, the choice of input variables to use in an ANN model is a fundamental problem that has significant impact on the prediction accuracy.

ANN can be a powerful tool if it is designed with the appropriate parameters. However, defining a topology with a suitable set of parameters, such as the number of input neurons, the number of hidden layers, hidden neurons, and weight values, to solve a complex problem can be considered an optimization problem.

The ANN topology can be fixed by a training process: The first step in this process is to select the best architecture according to the problem to solve. This step consists of defining the number of inputs, hidden and output neurons. The second step consists of finding optimum weight values that perform the ANN model. ANN architecture is generally chosen by experience, but in last years, some researches use metaheuristic algorithms as Particle Swarm Optimization (PSO) to search through a space of potential architectures, considering a fitness criterion.

In this paper, the main goal is to design a bankruptcy prediction model based on ANN, PSO and appropriate financial variables depending on data availability. So, the first issue is studying the contribution of variables selection models by comparing Multivariate Discriminant Analysis, Logistic Regression and Decision Trees. The second one is to define the optimum ANN topology by proposing a training process based on an improved PSO algorithm and

Simulated Annealing to find the optimal neural network topology.

The rest of paper is organized as follows: Section 2 presents the basic concepts used in this research; it defines models used in this research. In Section 3, the proposed methodology is detailed. Section 4 presents the empirical findings. Finally, in Section 5, we draw some conclusions and propose some further improvements.

## 2. Literature review

2.1 Bankruptcy prediction models

Many researches focused on designing predictive models to separate between failing and non-failing firms based on a set of financial variables. In literature, these models are categorized into two classes: statistical models and artificial techniques-based models.

The statistical models are the most known and the first used to predict failure. Early studies are made by Beaver [1] and Altman [2] and they are based on single and multi-discriminant analysis. Other statistical techniques are also used in failure prediction as logit and probit models [6] [7]. In statistical models, there are requirements concerning the variables selected as normal distribution, independency, high discriminate ability and complete observations availability [8].

With the appearance of artificial intelligence, many researchers have been interested to use these new techniques due to their ability to deal with imprecisely defined problems, incomplete and uncertain data in contrary with statistical models. For these qualities and because the topic of Firms' failure prediction has all these problematics, many studies in last decades use artificial intelligence-based models to predict failure especially neural networks. Odom and Sharda [9] were the first researchers that used neural networks in bankruptcy prediction and demonstrated its high predictive abilities compared with traditional method of bankruptcy prediction, which is multivariate discriminant analysis.

Many studies compared the ANN performance to other models: The study elaborated by [10] examined the predictive ability of four most commonly used financial distress prediction models: Multiple Discriminate Analysis, Logit, Probit, and Artificial Neural Networks. The results proof that the ANN approach demonstrates its advantage and achieves higher prediction accuracy where the data does not satisfy the assumptions of the statistical approach. Ruibin Geng [11] concluded that neural network is more accurate than other classifiers, such as Decision Trees and Support Vector Machines, as well as an ensemble of multiple classifiers combined using majority voting. Azayite and Achchab [12] studied the performance a hybrid discriminant analysis and artificial neural networks to predict firms' distress. The results exposed demonstrate that the proposed combined approach predict much accurate than the conventional neural network approach

2.2 Artificial Neural Network (ANN)

Inspired from neurobiology and defined as a branch of artificial intelligence, artificial neural network is able to build machines capable of learning and performing specific tasks such as classification, prediction or grouping. It is a collection of interconnected neurons that incrementally learn from their environment (data) to capture essential linear and nonlinear trends in a complex data, so that provides reliable predictions for new situations [3]

The first neuron model inspired from biological neuron was introduced by McCulloch and Pitts in 1943 [13]. They proof that formal neurons can perform logical functions. In 1949, parallel and connected neural networks models were introduced by the psychologist Donald Hebb. He then proposed many rules for updating the weights, the best known is the rule of Hebb [14].

In 1958, the psychologist Frank Rosenblatt developed the perceptron model [15]. it is a neural network that learns to recognize simple forms and to calculate some logical functions

The researches on artificial neural networks aroused a lot of interest until 1969, when two American scientists Minsky and Papert [14] demonstrated the limits of the perceptron particularly to resolve nonlinear problems. The interests to fully connected neural networks restarted from the 1980s with Rumelhart's publication [16] of the Back-Propagation algorithm. This learning algorithm optimize the multi-layered neural network parameters based on the propagation of the error towards the hidden layers.

Since then, the works and applications of neural networks in different fields have increased. In fact, it has been shown that Multi-layer perceptron network with a single hidden layer can approximate any function of $R^n$ in $R^m$ with an arbitrary precision [17].

The architecture and parameters of a neural network plays a fundamental role in its functionality and performance. A multilayer network is commonly used and it's an architecture composed with input layer, one or more hidden layers and output layer. The hidden layer defines the mapping between input neurons and output neurons. The relationships between layers are stored as weights of the connecting links. The layers are usually interconnected in a feed-forward way that means information moves in only one direction, forward, from the input layer, through the hidden layers, to the output layer with no cycles or loops in the network. To better understand the functioning of a neural network with propagation forward, we illustrate the topology of a network with a single hidden layer in Fig. 1. Moreover, in the literature, it has been proved that this architecture is the optimal one to solve a classification problem [18] which is the case of use for this article.

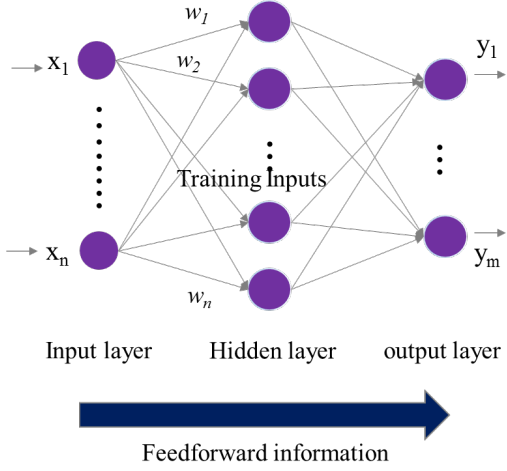

Fig. 1 : Artificial Neural Network

In the case of a network with a single hidden layer, each hidden neuron j = 1, ..., m, receives an input equal to the weighted sum of the inputs of the network then applied a transfer function f to transform input signal to an output signal defined as:

$$z_j = f\left(w_{j0} + \sum_{i=1}^{n} w_{ji} x_i\right) \quad (1)$$

with n and m are respectively the number of input and hidden neurons, $w_{ji}$ is the weight from the $i^{th}$ input neuron to the $j^{th}$ hidden neuron, $x_i$ is input variable i and $w_{j0}$ is a bias term.

The signals from the hidden neurons are then sent to output neurons through weighted connections in a similarity to what happens between input and hidden layer. As a result, the output nodes receive the sum of all weighted hidden neurons with an applied transfer function g depending on the desired output interval. The output $y_o$ of the output neuron o of the network is then formulated by:

$$y_o = g\left(b_{z0} + \sum_{i=1}^{m} \beta_{zj} \left(f\left(w_{j0} + \sum_{i=1}^{n} w_{ji} x_i\right)\right)\right) \quad (2)$$

With $\beta_{zj}$ is the weight from the $j^{th}$ hidden neuron to the $o^{th}$ output neuron and $b_{z0}$ is a bias term.

As mentioned before, the interest in the use of neural network in several domains is its characteristics to approximate any linear and non-linear function. However, the challenge is to find the topology of the neuron network and its weights value that approach the most the desired function. There can be seen as an optimization problem where we generally try to minimize a cost function based on the sum of the quadratic errors.

2.3 ANN optimization parameters

2.1.1 Input variables

After defining ANN, let's look at the first information needed for its construction which is input variables set. From these variables, mathematical models will be constructed, in our case, to separate two firms' classes healthy and bankrupt firms based on some financial variables. From learning examples, the ANN should learn to classify new features. Consequently, it is important to define variables that are the most relevant to do so.

As mentioned by F. Du jardin [5] a neural-network-based model for predicting bankruptcy performs significantly better when designed with appropriate variable selection techniques than when designed with methods commonly used in the financial literature. For this purpose, we use variables selection models to define appropriate financial ratios in one hand to select the most relevant variables, in other hand, to save money and effort for collecting and validating data especially for small size firms.

2.1.2 Architecture

As mentioned before, the architecture of neural network is composed by input, output and one or more hidden layers. So, other parameters that influence the NN performance and has to be considered in designing ANN is the number of neurons in each layer and the number of hidden layers. These parameters define the neural network behavior and depend to the problem to solve.

In literature, artificial neural network with one hidden layer is the best structure to use for classification problems [17] and this structure is used in this study.

Choosing the number of neurons in hidden layers could be a challenging task. If there are too many neurons, the number of possible computations that algorithm has to deal with increases. Otherwise, choosing few nodes in hidden layer can reduce the learning ability of the model [19]. So, it's very important to select the appropriate number of neurons that can maximize the network performance.

2.1.3 Learning algorithm

The process of setting the neural network parameters to mimic a specific behavior is called learning algorithm. It can be defined as a set of rules to find optimum values of weights and biases that maximize the neural network performance. There are various techniques that are used to find suitable values of ANN weights and biases depending on the type of learning. This type can be divided generally on two categories: supervised and unsupervised learning. Supervised learning is making the ANN learn from labeled examples. whereas unsupervised learning does not have labeled examples to guide learning, the weights of the neural network are modified according to specific criteria in order to discover regularities or clustering in observations.

In this article, we are more interested to optimize an ANN to predict failure and separate between healthy and bankrupt firms when it is trained to assign the correct target classes to a set of labeled firms' variables. In this supervised learning, the class labels are pre-determined and provided in the training phase, a learning algorithm tries to estimate the dependencies between inputs and targets given by adjusting the connection weights until an error function, like mean squared error, is minimized.

There are different learning methods. We can classify them on two main groups:

• The first group based on steepest descent techniques. It includes methods as gradient descent, Levenberg-Marquardt, Backpropagation and other variants. Some of these algorithms require a considerable amount of time and memory. The most used algorithm is the Backpropagation algorithm. It is a powerful algorithm for calculating the gradient of neural networks but has the limits especially the problem of local minimum.

• The second group includes methods based on the evolution of living species and which includes, among others, genetic algorithms, swarm algorithms, simulated annealing …

These algorithms have been designed to search for the global optimum.

2.1.4  Transfer function

One of parameters that should be fixed before training a neural network is the transfer function. The choice of using an activation function depends on the use case: for example, the binary function is adapted to the problems of organization or distribution. Whereas continues and differentiable functions as linear, gradient and sigmoid functions are used to approximate continues functions.

It is also interesting to note that particularly the sigmoid transfer function is the most used. It is because it combines nearly linear behavior, curvilinear behavior, and nearly constant behavior, depending on the value of the input [19]. This property allows artificial neural network to adapt both linear and non-linear problems. The sigmoid function is formulated as:

$$f(x) = \frac{1}{(1 + e^{-x})} \qquad (3)$$

It takes any real-valued input and returns an output bounded between zero and 1. This function is used in this study as transfer function.

2.4 Particle Swarm Optimization (PSO)

Developed by a social psychologist J. Kennedy and an electrical engineer R. Eberhart in 1995 [20] and described as evolutionary computation method, PSO is one of the swarm intelligence algorithms that are inspired by the natural process of social behavior of organisms as birds flocking and used an optimization technique in many research areas.

Animals living in swarms can travel long distances for migration or food search and must, for this purpose,

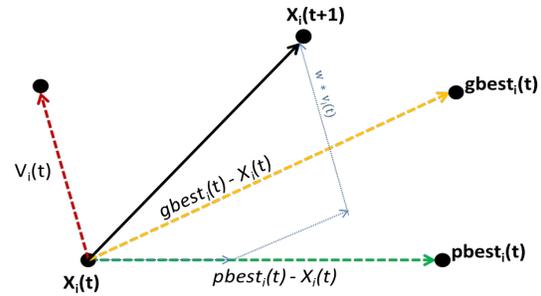

Fig. 2 : PSO particle movement

optimize their movements in terms of time and energy expended and converge towards the food source. The movement of these animals is complex, each member of the group determines its movement according to its experience as well as that of the other members. The PSO is inspired by this behavior in order to approximate solutions for problems and it is used to optimize a continuous function in a data space.

PSO algorithm, as detailed in Algorithm 1, is based on a set of individuals called particles, initially randomly arranged, that move in the solution space in search of the best solution. Each position of the particle represents a potential solution to the problem. The movement of each particle in the search space, influenced by the move of the other particles of the neighborhood, follows specific rules: Each particle has a memory that allows it to memorize the best point by which it has already passed, and it tends to return to that point. Each particle is informed of the best point encountered in its neighborhood and it tends to go to that point.

Algorithm 1: PSO algorithm

```
For each particle i in swarm S
Initialize particle i
End
While stopping condition is false do
        For each particle i of the swarm do
                Calculate the fitness f(xi(t));
                If the fitness value>pbest
Set current value as the new pbest
                End
        End for
Choose the best fitness value of all the particles as gbest
For each particle i of the swarm do
Update particle velocity according equation (1)
Update particle position according equation (2)
        End for
End while
```

To apply PSO, we need to define the search space composed by particles and a fitness function to optimize. The system is then initialized with a population of random solutions (particles) where each particle has a position value representing a possible solution data, a velocity value indicating how much the Data can be changed and a personal best value (pBest) that indicates the best solution reached by the particle.

The particle velocity is influenced by three components represented in Figure 2 and detailed as:

*Physical component:* The particle tends to follow its own path through

$$\text{Physical vector} = w * \vec{v}(t)$$

*Cognitive component*: It tends to return to the best position by which it has already passed

$$\text{Cognitive vector} = c_1 * r_1 * (\overrightarrow{pBest}(t) - \vec{x}(t))$$

*Social component*: It tends to move towards the best position already reached by its neighbors

$$\text{Social vector} = c_2 * r_2 * (\overrightarrow{gBest}(t) - \vec{x}(t))$$

With w is an inertia weight, v velocity of the particle in the iteration t, $c_1$ and $c_2$ are respectively cognitive and social parameters, $r_1$ and $r_2$ random numbers between 0 and 1, p the particle best position (pBest), x the particle current position and g the swarm best position (gBest).

The equation to update velocity of the particle is then:

$$\vec{v}(t+1) = (w * \vec{v}(t)) + (c_1 * r_1 * ((\overrightarrow{pBest})(t) - \vec{x}(t)) + (c_2 * r_2 * ((\overrightarrow{gBest})(t) - \vec{x}(t))) \quad (2)$$

The velocity v is limited to the range [$V_{max}$, $V_{min}$]. Updating velocity in this way enables the particle to search for its best individual position pBest and the particle position is computed as:

$$\vec{x}(t+1) = \vec{x}(t) + \vec{v}(t+1) \quad (3)$$

## 2.5 Simulated Annealing

Simulated annealing (SA) is a stochastic optimization algorithm that was introduced by [18]. As its name denotes, simulated annealing is inspired by a physical metal annealing process in which it is heated very strongly, and then gradually cooled to form a perfect crystalline structure with minimal energy. SA is considered as an optimization algorithm that simulates an analogy between the way a metal cools into a minimum energy crystalline structure and the search for a minimum in a general optimization problem. Indeed, in the annealing process of the metal, the physical states of the material, the energy of a state and the temperature respectively correspond by analogy to the solutions of a problem, the cost of a solution and a control parameter.

In the search process, SA accepts not only the best solution but also the worst solution with a probability. At the beginning of the process, the temperature is high and the probability of accepting a worse solution is also high, but as the temperature decreases, the probability of acceptance of the less good solution gradually approaches zero.

The principle of SA is to explore the search space in an iterative way. Thus, the algorithm starts from a random initial solution noted S0 which corresponds to energy E0 and a high initial temperature T0. Then, conditioned iterations follow where for each iteration, the objective function E is calculated and compared with the previous value. If the new value of the objective function is better, the new solution is automatically accepted; otherwise it can be accepted with a probability P which depends on the current temperature T and the difference of the objective function ΔE. The probability P is:

$$P = e^{-\Delta E/T} \quad (4)$$

At a very high temperature, this probability of acceptance is very high which results in a systematic acceptance of any solution. This phase corresponds to a random local walk in the search zone. On the contrary, at a low temperature the probability decreases which means that there is little chance for the acceptance of a less good solution. This phase corresponds to a local search.

This mechanism allows exploring a new area of solutions, better or worse in order to leave the local minima.

## 3. Research methodology

Considering the good capability of optimization of PSO, we are interested in this study to analyze the use of the Particle Swarm Optimization algorithm and propose a new hybrid model to optimize a neural network topology. So, the proposed methodology has the objective to design the optimum ANN topology that deals with noisy and missing Data to identify the firms' health with a high accuracy. This approach detailed in Figure 3 is presented and detailed below.

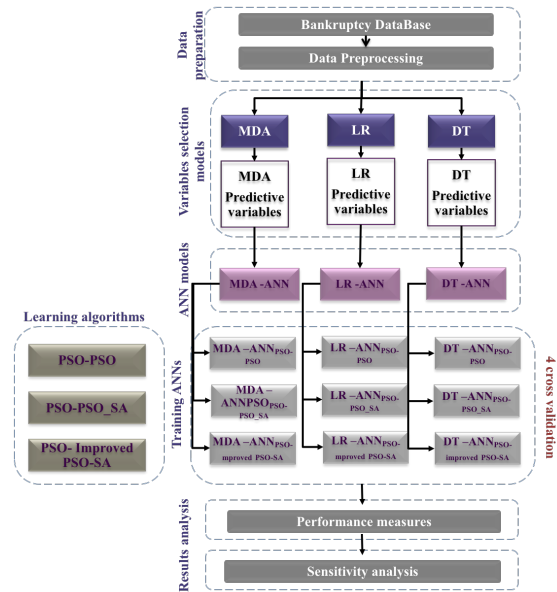

Fig. 3 : The proposed methodology

## 3.1 Financial variables

Before looking how can we find the optimum ANN architecture, let's look at the first information needed for its definition which are the input variables. From these variables, mathematical models will be constructed, in our case, to predict business failure. So, one of the challenges that can affect the performance of a failure distress model is the definition of the appropriate financial ratios. There is no universally agreed financial ratios list to use in financial performance prediction models and suppliers and customers' payment delays are not commonly used as discriminant ones. For instance, Altman [2] used 5 ratios, Liang & Wu [19] used 7 ratios, [4] used 35 ratios.

Table 1 : Initial financial ratios list

| Indicators Group | Indicator | Description |
|---|---|---|
| **Liability** | Debt Ratio | Total debt / Owners' Equity |
| | Debt capability | financial debt / Owners' Equity |
| | Intercompany debt | Accounts payable and receivable / total liabilities |
| | Financial debt ratio | Financial debt and cash flow / Owners' Equity |
| | Repayment capacity | Long and medium term debt / Cash flow |
| | Short term debt-to-Total debt | Short-term debt / total liabilities |
| | Long term debt-to-Total debt | Long and medium term debt/ total liabilities |
| | Weight of interest charges | Interest expenses / EBITDA |
| | Financial autonomy | Owners' Equity / permanent capitals |
| | Debt coverage by equity | Owners' Equity / financial debt, cash flow and Accounts payable and receivable |
| | Permanent capitals-to-Fixed assets | Permanent capitals / Fixed assets |
| | Current assets-to-Short term debt | Current assets / Short-term debt |
| **Profitability** | Profitability ratio | Retained earnings / totat assets |
| | Return on Equity | Net Income/ Owners' Equity |
| | Return on Assets | Net Income / Total Assets |
| | Margin rate | EBITDA / Earning value |
| | EBITDA Margin | Earnings Before Interest, Tax, Depreciation, and Amortization / Sales |
| | Net Profit Margin | Net Income/Sales |
| **Operating ratios** | Customers' payment delays | accounts receivable / turnover * 360 |
| | Suppliers' reimbursement delays | Suppliers and related accounts / Purchases *360 |
| | Stock rotation | Total inventories / Purchases *360 |
| | Finished goods awaiting to be Rotation | Finished goods awaiting to be sold/ Turnover *360 |
| | Rotation of goods | Cost of a merchandises / Purchases *360 |
| | Salaries Payable rate | Salaries payable / Earning value |
| | Taxes rate | Taxes and duties / Earning value |
| | Financial charges rate | Financial charges / Earning value |
| **Liquidity** | Working capital-to-total assets | Working capital /Total assets |
| | Quick Ratio | (Current Assets – Inventory) / Current liabilities |
| | Liquidity Ratio | Current Assets / Current liabilities |
| | Cash Ratio | Cash and Cash Equivalents / Current liabilities |
| **Turnover** | Receivable Turnover Rate | Sales / Accounts Receivable |
| | Net Working Capital Turnover Rate | Sales / (Current Assets – Current Liabilities) |
| | Asset Turnover Rate | Sales / Total Assets |
| | Fixed Asset Turnover Rate | Sales/ Fixed Assets |
| | Current Assets Turnover Rate | Sales / Current Assets |
| **Productivity** | Productivity ratio | Earnings before interest and taxes / Total assets |
| **Balance sheet raws** | | Turnover |
| | | total assets |
| | | Fixed assets |

So, depending on data availability, we used 39 financial ratios defined in Table 1 and classified into seven categories such as profitability, liquidity and productivity and we also added an operating ratios group that contains variables that are not commonly used containing Customers' payment delays, Suppliers' reimbursement delays and Stock rotation.

### 3.2 Data preprocessing

Nowadays data storage capacity is no longer a problem, there is more and more data available. However, their quality is to be discussed and we a faced to be dealing with missing, noisy or redundant data issues.

Experimental studies have shown that many algorithms display weaknesses because of
redundant or poor quality introduced variables. For this reason, two steps are considered to prepare Data for building a predictive model: the dimensionality reduction through a variable selection model and data preprocessing.

Data preprocessing is a crucial step that is often overlooked and rarely addressed in classification issues. This step includes Data preparation and normalization to perform reduction or classification tasks. Outliers in the learning database can affect the performance of any model including neural networks and can be reduced by normalizing variables.

### 3.3 Variables selection models

In binary classification researches, it is important to define which variables are the most relevant to separate two classes. it is also common to have difficulties to collect reliable data and calculate financial ratios especially for small size firms. This is the reason why in this case, it's important to define the most important variables that can provide information to predict a risk failure so to save money and effort for collecting and validating data.

To do so, the reduction of dimensionality is one of the solutions to consider in setting up a prediction model. This will not only reduce the computational complexity but also improve the performance of classification algorithms including neural networks.

In the same context, F. Du jardin [5] has shown that a neural-network-based model for predicting bankruptcy performs significantly better when designed with appropriate variable selection techniques than when designed with methods commonly used in the financial literature.

For this reason, to identify the significant and the minimum set of variables that can discriminate bankruptcy and healthy firms and in order to choose the appropriate variable selection model that maximizes the ANN performance, we compare three models classified into two fields of classification techniques, statistic methods and artificial intelligence techniques.

Statistical method used is chosen for its popularity in the financial literature and the predominate one is Multivariate Discriminant Analysis (MDA) and Linear Regression (LR).
MDA is used to find a linear combination of features that characterizes or separates two or more classes of objects or events and, more commonly, for dimensionality reduction before later classification. In this research, we used a stepwise method applying Wilk's Lambda to select sequentially significant variables from the set of variables.

LR is an econometric method that addresses the problem of regression when the dependent variable is binary (or discrete in the most general case). Indeed, the ordinary least squares method usually used to regress a continuous variable by a set of explanatory variables is no longer appropriate in so far as the variable to be explained has no more value in the space R but rather in a finite set ((0,1) for example in the binary case).

As a popular intelligence technique to define appropriate variables, we select Decision Trees (DT). it's a tool used to discover the relationship between variables. Its main use is for segmentation or tree growing. Based on adjusted significance testing, it can be used for prediction, regression analysis, classification, and for detection of interaction between variables.

### 3.4 ANN architecture

In this study, we use ANN based model to predict failure. As mentioned before, the topology of ANN plays a fundamental role in its functionality and performance. So, we are interested to define the best architecture that can discriminate between healthy and bankrupt firms depending on their financial statements.

To design an ANN, there are parameters that need to be defined as number of inputs, number of hidden layers, number of hidden neurons.

In literature, artificial neural network with one hidden layer is the best structure to use for classification problems [17]. The number of inputs depends on variables selection techniques used in this study. Each model selects appropriate variables set that is used as input variables for the ANN. So, three hybrid neural networks are compared in this study: LR-ANN, MDA-ANN and DT-ANN.

### 3.5 Training ANNs

After defining the number of input neurons, we need to find the optimum parameters as number of hidden neurons and the weight values to fix the ANN topology. To do so, we need a training process. It is generally made in two phases: The first one is the selection of appropriate neural network architecture by defining number of hidden neurons and it is generally chosen by experience, but in last years, some researches use metaheuristic algorithms as Genetic Algorithms (GA) or Particle Swarm Optimization (PSO) to search an optimal solution through a space of potential architectures, considering a fitness criterion. The second phase is adjusting neurons weights until finding the optimum ones that perform the ANN. In general, initial weights and bias are set randomly and there are adjusted during the training phase by a learning algorithm.

Considering the good capability of optimization, we are interested in this work, on particle swarm optimization (PSO) and Simulated Annealing (SA) to use as alternative to escape local minima and to search for a global optimum that perform the bankruptcy prediction neural network model.

In this work we compare three learning algorithms based on PSO: PSO-PSO, PSO-PSO_SA and PSO-ImprovedPSO_SA.

3.5.1 PSO-PSO

The first algorithm is inspired by Zhang and Shao's methodology [20] and it's based on two loops of PSO: the first loop of PSO algorithm is used as number of hidden neurons optimization process interleaved with a second optimizing weight values PSO. The corresponded algorithm is presented in Algorithm 2.

3.5.2 PSO-PSO_SA

the second algorithm compared in this study is PSO-PSO_SA. it is based on a hybrid PSO and SA to accelerate the convergence of the PSO.

In the inner PSO and in order to evolve the convergence of the PSO specially to escape local minima and find optimum weights, we combine it with Simulated Annealing. The flow chart of

Algorithm 3 : PSO-PSO algorithm

```
For each particle in swarm S- the population of architectures A
    Initialize particle
    Create network with a random n hidden neurons
End
While stopping condition 1 is false do
    For each particle A_i of the swarm A do
        //execute PSOalgorithm (A_i weights)
            Generate particles swarms Bj with random weights matrix as position.
            While stopping condition 2 is false do
                For each particle j of the swarm do
                    Calculate the fitness f(x_j(t));
                    If the fitness value >pBest(B)
        Set current value as the new pBest(B)
                    End
                End for
        Choose the best fitness value of all the particles in B swarm as gbest(B)
        For each particle j of the B swarm do
            Update particle j velocity according equation (2)
            Update particle j position according equation (3)
                End for
            End while
        Fitness value = gBest(B);
            If the fitness value >pBest(A)
        Set current value as the new pBest(A)
                End
            End for
    Choose the best fitness value of all the particles as gBest
End for
Foreach particle Ai of the swarm A do
        Update particle velocity according equation (2)
        Update particle position according equation (3)
        Update the A_i.net to the new architecture represented by A_i;
End for
End while
```

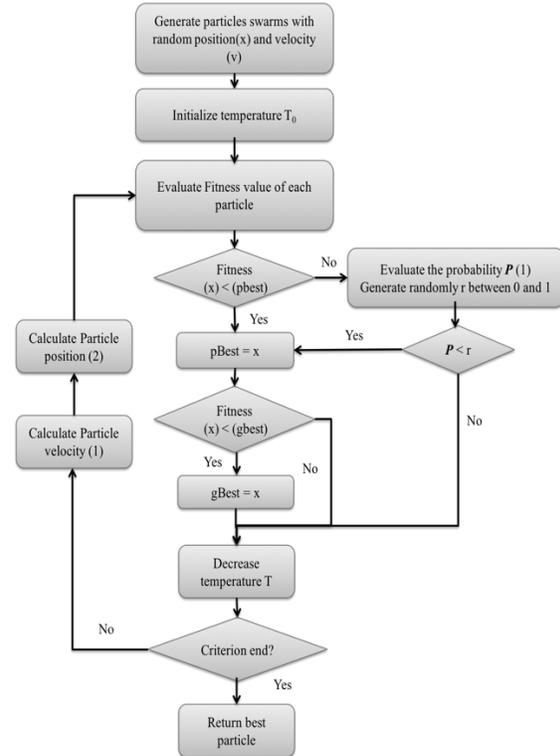

Fig. 4 : Flow chart PSO-SA algorithm

PSO_SA is detailed in Figure 4. So, the learning algorithm PSO-PSO-PSO_SA is based on PSO interleaved with a proposed PSO_SA. The outer PSO is used to optimize the number of hidden nodes whereas the inner hybrid PSO_SA is used to optimize the weights for each of architectures given in the outer PSO.

3.5.3 PSO-improvedPSO_SA

The last learning algorithm that is compared in this research is based on PSO interleaved with the proposed hybrid Improved PSO and simulated annealing. The improved PSO presented in this study is based on the hypothesis that learning is a process based not only on good experiences but also on bad experiences. For this reason, we introduce in the Velocity equation (2) of the standard PSO, the pWorse value that corresponds to the last bad point reached by the particle. The Velocity of the new PSO algorithms is then calculated as:

Algorithm 4 : PSO-improvedPSO-SA algorithm

```
For each particle in swarm S- the population of architectures A
Initialize particle
Create network with a random n hidden neurons
End
While stopping condition 1 is false do
   For each particle A_i of the swarm A do
      //execute PSO algorithm (A_i weights)
      Generate particles swarms Bj with random weights matrix as position.
         While stopping condition 2 is false do
         For each particle j of the swarm do
         Calculate the fitness f(x_j(t));
         If the fitness value >pBest(j)  //case of best solution
Set current value as the new pBest(j)
         Else   // apply SA in worse solutions
         Compute ΔE = fitness f(x_j(t)) - the fitness f(x_j(t-1))
                Probability = exp(-ΔE/KT)
         If Probability > random number between 1 and 0
            Accept the worse solution set current value as the new pWorse(j)
               End
         End for
         t = frac * t;  // frac < 1// Lower the temperature for next cycle
Choose the best fitness value of all the particles in B swarm as gbest(B)
For each particle j of the B swarm do
Update particle j velocity according equation (4)
Update particle j position according equation (3)
End for
End while
Fitness value = gBest(B);
   If the fitness value >pBest(A)
      Set current value as the new pBest(A)
         End
         End for
Choose the best fitness value of all the particles as gBest
End for
For each particle Ai of the swarm A do
      Update particle velocity according equation (3)
      Update particle position according equation (2)
      Update the Ai.net to the new architecture represented by Ai;
   End for
   End while
```

$$\vec{v}(t+1) = (w * \vec{v}(t)) + (c_1 * r_1 * (\vec{p}(t) - \vec{x}(t)) - (c_3 * r_3 * (\overline{pWorse}(t) - \vec{x}(t)) + (c_2 * r_2 * (\vec{g}(t) - \vec{x}(t)) \quad (4)$$

The inclusion of the last worse point reached in the behavior of the particle gives additional information in the cognitive component. Bad experience vector allows particle to move away from its last worse point and try to search for a best solution. Also, in this algorithm the w inertia value is linearly changed for each iteration.

The improved PSO is combined with SA to find optimum weights and to avoid the problem of local minima. SA is used in this context, as trial way to explore new solutions in the defined search space, either better or worst. As the purpose of SA is accepting a worse solution according to a probability that gradually decreases as the temperature decreases to approach zero, this technique allows jumping out from local minima. The proposed PSO-ImprovedPSO-SA pseudo code is detailed in Algorithm 3. The fitness function used to evaluate the performance of each particle is the Mean Squared Error

### 3.6 Cross validation technique

Any bias due to Data set selected could have a detrimental impact on determining NN architecture and its parameters. A cross validation technique is introduced to minimize the bias effect.
In this study, a 4-fold cross validation technique is used to train and test the hybrid ANN models to avoid over fitting. For each model, the data set is randomly split into 4 mutually exclusive subsets of approximately equal sizes. The model is trained first and tested 4 times. Each time, the model is trained on 3 folds and tested on the remaining 1-fold. The overall accuracy of a model is evaluated by averaging the 4 individual accuracy measures

### 3.7 Performance evaluation

To compare results, we use this list of measures:
*Overall accuracy:* it is defined as the percentage of records that are correctly predicted by the model. According to confusion matrix in Table 2 the formula is:
$$Accuracy = \frac{TP + TN}{TP + TN + FP + FN}$$

*Precision:* it is defined as the ratio of the number of True Positive (correctly predicted cases) to the sum of the True Positive and the False Positive.
*Recall:* it is known as the Sensitivity or True Positive rate. It is defined as the ratio of the True Positive (the number of correctly predicted cases) to the sum of the True Positive and the False Negative.
*Specificity:* It is defined as the ratio of the number of the True Negative to the sum of the True Negative and the False Positive.

Table 2 : Confusion matrix

| | Predicted | |
|---|---|---|
| Actual | *Unsuccessful* | *Successful* |
| *Unsuccessful* | True Negative | False Positive |
| *Successful* | False Negative | True Positive |

*F-Measure:* F-measures take the harmonic mean of the Precision and Recall Performance measures

$$F - Measure = 2 \cdot \frac{Precision \cdot Recall}{Precision + Recall}$$

## 4. Experiments and evaluation

As mentioned before, the main goal of this research is to design a bankruptcy prediction neural network mainly based on Particle Swarm Optimization. To do so, the first step is to collect and select the appropriate set of variables that can be used to build neural network model.

4.1 Data and variables

For the experiment, the database used contains annual financial statements data for a sample of Moroccan firms. Different samples are collected over the period from 2010 to 2013. We call a company bankrupt if, according to companies' public registry, it has removed (the final stage of bankruptcy). Three firms' sectors are selected to reduce the influence of the firms' activity: retail, services and manufacturing. Data collected contains 690 companies and are almost balanced. A binary variable is created with two values (0 if it is failed and 1 if it is healthy).

Data collected represent the calculated financial ratios detailed in 3.1 paragraph. In this step it is common to have difficulties to collect reliable data and calculate financial ratios especially for small size firms, as they do not present complete balance sheet. This is why, in the initial database, we have the missing data constraint and one of the challenges of the models used in this research is to predict bankruptcy and give a good accuracy based on a minimum set of variables.

To reduce the effect of outliers and to perform reduction and classification tasks, Data used in this paper, is normalized.

4.2 Results

In the context of missing data, reducing variables and analyzing the most important ones is very interesting. So, the first step to build a bankruptcy prediction model is to define the relevant variables that separate between bankrupt and healthy firms through the use of variable selection techniques. The models compared in this paper are those most commonly used in the literature: Discriminant Analysis, Logistic Regression and Decision Tree (union of CHAID and CRT). The variables selected by each technique is summarized in Table 3

As detailed in Table 3, each model has selected its own variables depending on their discrimination ability. Three sets of variables have been constructed; 8, 13 and 16 variables have been chosen respectively by MDA, LR and DT.

Each set of variables will serve as input variables layer in an ANN model. Consequently, three hybrid neural network models are considered: MDA-ANN, LR-ANN and DT-ANN.

Table 3 : Variables selection models results

| *Variable selection model* | *Number of selected indicators* | *Selected financial indicators* | |
|---|---|---|---|
| Multivariate discriminant analysis | 8 | total assets | Short term debt-to-Total debt |
| | | Working capital-to-total assets | Salaries Payable rate |
| | | Profitability ratio | Taxes rate |
| | | Productivity ratio | Liquidity Ratio |
| Logistic Regression | 13 | total assets | Financial debt ratio |
| | | Fixed assets | Customers' payment delays |
| | | Working capital-to-total assets | Finished goods awaiting to be Rotation |
| | | Profitability ratio | Quick Ratio |
| | | Productivity ratio | Liquidity Ratio |
| | | Asset Turnover Rate | Cash Ratio |
| | | EBITDA Margin | |
| Decision Tree | 16 | total assets | Profitability ratio |
| | | Return on Assets | Permanent capitals-to-Fixed assets |
| | | Suppliers' reimbursement delays | Net Working Capital Turnover Rate |
| | | Quick Ratio | Turnover |
| | | productivity ratio | Debt Ratio |
| | | Taxes rate | Stock rotation |
| | | Turnover | Profitability ratio |
| | | Fixed assets | EBITDA Margin |
| | | Customers' payment delays | Intercompany debt |

In order to design the best topology for each hybrid neural network and as mentioned before we compare three learning algorithms based on PSO: PSO-PSO, PSO-PSO_SA and PSO-ImprovedPSO_SA.

To use the algorithms based on two loops of PSO as presented before, we used the parameters below:

- PSO for architectures optimization

– swarm size (s = 20) and stop criteria (15 iterations)
– Search space limit [7, 30]
– inertia factor ($w_n = 0.9\, w_{n-1}$, $w_0 = 0.8$)
– Acceleration factors ($c_1 = c_2 = 1.4960$)

- PSO for weights optimization

– swarm size (s = 50) and stop criteria (2000 iterations)
– Search space limit [−2.0, 2.0]
– inertia factor ($w_n = 0.9\, w_{n-1}$, $w_0 = 0.8$)
– Acceleration factors ($c_1 = c_2 = 1.4960$)

After training these three algorithms applied to the three hybrid ANN, the topology that gives the best accuracy is 16-23-1 (16 input neurons, 23 hidden neurons and one output neuron) and the results obtained are summarized in Table 4.

The hybrid DT-ANN trained with the proposed algorithm *PSO-improvedPSO-SA* displays the best overall accuracy (84.2%). This result means that variables introduced in this model based on Decision tree as variable selection model have a good power to discriminate bankrupt and non-bankrupt firms. Also, we can observe that for each hybrid neural network, introducing Simulated Annealing ameliorates the performance of PSO: *PSO-PSO_SA* and *PSO-improvedPSO-SA* give better results than *PSO-PSO*. For the three hybrid models, the proposed algorithm PSO-improvedPSO-SA gives the best results. These conclusion means that the improved PSO presented in this study, based on introducing bad experiences in the PSO particles velocity can ameliorate the performance of the learning algorithm and that learning is not based on good experiences but also on bad experiences.

In many studies, the models are compared through the overall accuracy to judge the performance. They do not consider per-class performance which gives important information about a model specially to select variables that discriminate bankrupt to healthy firms. The cost of bad prediction for a bankrupt firm is not the same as for healthy firm, especially for banks, investors, leaners. Following this measure, it shows that the hybrid DT-ANN even with its best overall accuracy, present the less rate of good classification of bankrupt firms (between 71.1% and 73.5%) contrary to non-bankrupt firms (between 91.2% and 92.2%). The MDA-ANN model gives the best bankrupt classification rate (between 82.7% and 83.4%) and the worst non-bankrupt firms' rate (between 44% and 48.3%). These results mean that the variables selected by statistical models especially MDA allow to give information only about bankrupt firms and DT variables almost separate between the two classes.

In order to study the convergence of each algorithm used to train the hybrid Artificial Neural Networks developed in this paper, we plot the error cost calculated per iteration in the inner loop of PSO for the best particle observed by the algorithm.

For the hybrid DT-ANN architecture, Figure 5 present the evolution of the cost error of the best particle trained through the three algorithms compared PSO-improvedPSO_SA, PSO-PSO_SA and PSO-PSO. Figure 6 and Figure 7 present the best particle convergence for the hybrid LR and ANN and the hybrid MDA and ANN respectively.

Based on these figures, we can observe that the three best particles for the different hybrid ANN find a local minimum in less than 30 iterations especially PSO-PSO that is the first algorithm that find a minimum error cost in less than 30 iterations. But these algorithm in the three figures stagnate in these point that means that the particle is caught in a local minimum. In contrary, the two other algorithms based on Simulated Annealing continue their search for a global minimum. These results confirm the hypothesis that introducing SA as trial way to explore new solutions in the defined search space allows the particle to find optimum weights and avoids the problem of local minima. In the three hybrid ANN developed in this study, the training algorithm PSO-improvedPSO_SA presents the best algorithm that allows to find appropriate ANN weights which minimize the error cost.

Table 4 : Classification results of Hybrid Neural network models

| Model | Overall accuracy | Precision | Sensitivity | Specificity | F-measure | Bankrupt | Non-bankrupt |
|---|---|---|---|---|---|---|---|
| DT-ANN | | | | | | | |
| DT-ANN$_{PSO-PSO}$ | 83,0% | 91,2% | 81,5% | 85,8% | 86,1% | 72,1% | 91,2% |
| DT-ANN$_{PSO-PSO\_SA}$ | 83,2% | 92,2% | 81,1% | 87,1% | 86,3% | 71,1% | 92,2% |
| DT-ANN$_{PSO-improvedPSO\_SA}$ | **84,2%** | **92,2%** | **82,4%** | **87,4%** | **87,0%** | **73,5%** | **92,2%** |
| LR-ANN | | | | | | | |
| LR-ANN$_{PSO-PSO}$ | 72,5% | 70,1% | 73,1% | 72,0% | 71,6% | 74,9% | 70,1% |
| LR-ANN$_{PSO-PSO\_SA}$ | 73,8% | 73,7% | 73,3% | 74,2% | 73,5% | 73,9% | 73,7% |
| LR-ANN$_{SO-improvedPSO\_SA}$ | 75,0% | 72,9% | 75,6% | 74,5% | 74,2% | 77,0% | 72,9% |
| MDA-ANN | | | | | | | |
| MDA-ANN$_{PSO-PSO}$ | 62,9% | 44,0% | 72,6% | 58,6% | 54,8% | 82,7% | 44,0% |
| MDA-ANN $_{PSO-PSO\_SA}$ | 65,0% | 47,5% | 74,9% | 60,3% | 58,1% | 83,4% | 47,5% |
| MDA-ANN $_{PSO-improvedPSO\_SA}$ | 65,4% | 48,3% | 75,1% | 60,6% | 58,8% | 83,3% | 48,3% |

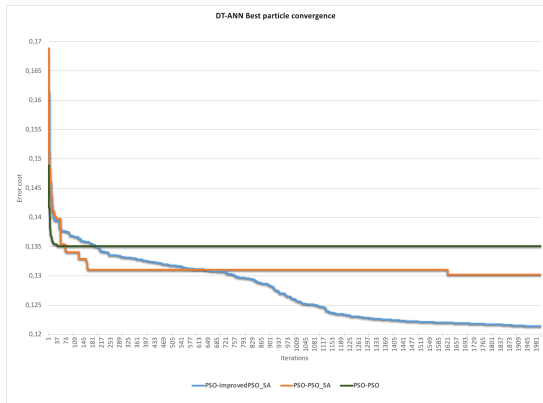

Figure 5 : DT-ANN best particle convergence

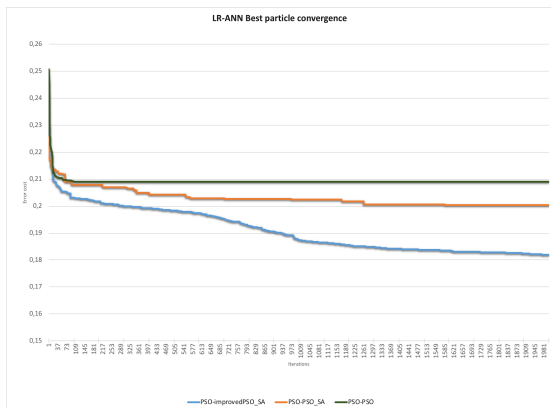

Figure 6 : LR-ANN best particle convergence

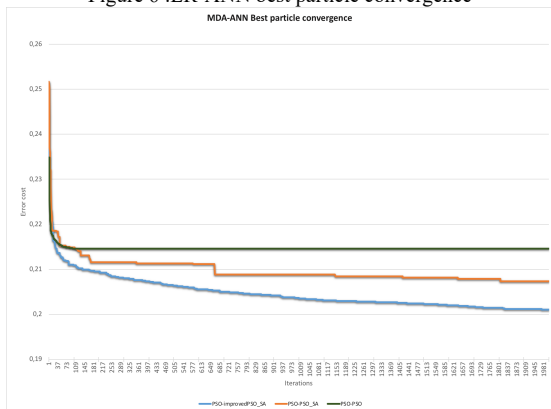

Figure 7 : MDA-ANN best particle convergence

## 4. Conclusions

In this paper, we propose a hybrid ANN to predict failure based on Particle Swarm Optimization and variables selection techniques.

The methodology proposed studied the contribution of variables selection models by comparing Multivariate Discriminant Analysis, Logistic Regression and Decision Trees. The results show a high performance of Decision Trees as variables selection models for ANN to discriminate between Bankrupt and non-bankrupt firms.

Furthermore, we propose a training algorithm *PSO-improvedPSO_SA* to find the optimum topology. The training algorithm is based on the use of Simulated Annealing that allows jumping out from local minima and the hypothesis that learning is a process based not only on good experiences but also on bad experiences. The proposed algorithm gives high performances especially when applied to the hybrid DT and ANN model. This model is a good early warning system to use by investors and creditors.

**Fatima Zahra Azayite** A computer science engineer, a data analyst and PhD student in National School for Computer Science and Systems analysis. Her principle research domains are machine learning, business intelligence and optimization algorithms. She currently works in a central bank and has ten years' experience in Data analysis, Business Intelligence and statistics.

**Said Achchab** A graduate of the Mohammedia School of Engineering with a PhD in Applied Mathematics and a university qualification in Business Intelligence, Saïd Achchab also followed the senior management cycle of ENCG Settat. He is currently Professor of Quantitative Finance, Business Intelligence and Change Management at ENSIAS as well as Director of the Continuing Education Center of this same school.